\documentclass[journal=jacsat,manuscript=article]{achemso}

\usepackage[version=3]{mhchem} 
\usepackage[T1]{fontenc}       
\usepackage{graphicx,color}

\author{Jinwoong Hwang}
\affiliation{Department of Physics, Pusan National University, Busan 46241, Korea}

\author{Kyoo Kim}
\affiliation{Max Planck-POSTECH/Hsinchu Center for Complex Phase Materials, Pohang University of Science and Technology, Pohang 37673, Korea}

\author{Hyejin Ryu}
\affiliation{Max Planck-POSTECH/Hsinchu Center for Complex Phase Materials, Pohang University of Science and Technology, Pohang 37673, Korea}
\alsoaffiliation{Advanced Light Source, Lawrence Berkeley National Laboratory, Berkeley, California 94720, USA}
\alsoaffiliation{Center for Spintronics, Korea Institute of Science and Technology, Seoul 02792, Korea}

\author{Jingul Kim}
\affiliation{Department of Physics, Pohang University of Science and Technology, Pohang 37673, Korea}

\author{Ji-Eun Lee}
\affiliation{Department of Physics, Pusan National University, Busan 46241, Korea}

\author{Sooran Kim}
\affiliation{Max Planck-POSTECH/Hsinchu Center for Complex Phase Materials, Pohang University of Science and Technology, Pohang 37673, Korea}
\alsoaffiliation{Department of Physics, Pohang University of Science and Technology, Pohang 37673, Korea}

\author{Minhee Kang}
\affiliation{Department of Physics, Pusan National University, Busan 46241, Korea}

\author{B. -G. Park}
\affiliation{Pohang Accelerator Laboratory, Pohang University of Science and Technology, Pohang 37673, Korea}

\author{A. Lanzara}
\affiliation{Materials Sciences Division, Lawrence Berkeley National Laboratory, Berkeley, California 94720, USA}
\alsoaffiliation{Department of Physics, University of California, Berkeley, CA 94720, USA}

\author{J. W. Chung}
\affiliation{Department of Physics, Pohang University of Science and Technology, Pohang 37673, Korea}
\alsoaffiliation{Department of Physics and Photon Science, Gwangju Institute of Science and Technology, Gwangju 61005, Korea}

\author{S. -K. Mo}
\affiliation{Advanced Light Source, Lawrence Berkeley National Laboratory, Berkeley, California 94720, USA}

\author{J. Denlinger}
\affiliation{Advanced Light Source, Lawrence Berkeley National Laboratory, Berkeley, California 94720, USA}

\author{B. I. Min}
\affiliation{Department of Physics, Pohang University of Science and Technology, Pohang 37673, Korea}

\author{Choongyu Hwang}
\affiliation{Department of Physics, Pusan National University, Busan 46241, Korea}
\email{ckhwang@pusan.ac.kr}

\title
  {Emergence of Kondo resonance in graphene intercalated with cerium}
  
\begin{document}

\newpage

\begin{abstract}

The interaction between a magnetic impurity, such as cerium (Ce) atom, and surrounding electrons has been one of the core problems in understanding many-body interaction in solid and its relation to magnetism. Kondo effect, the formation of a new resonant ground state with quenched magnetic moment, provides a general framework to describe many-body interaction in the presence of magnetic impurity. In this Letter, a combined study of angle-resolved photoemission (ARPES) and dynamic mean-field theory (DMFT) on Ce-intercalated graphene shows that Ce-induced localized states near Fermi energy, $E_{\rm F}$, hybridized with the graphene $\pi$ band, exhibit gradual increase in spectral weight upon decreasing temperature. The observed temperature dependence follows the expectations from the Kondo picture in the weak coupling limit. Our results provide a novel insight how Kondo physics emerges in the sea of two-dimensional Dirac electrons.

\end{abstract}

The Kondo effect arises from the interaction between the local moment of magnetic impurities and the spins of conduction electrons in surrounding non-magnetic metallic host as shown in Fig.~1~\cite{Kondo,KL}. The presence of the impurity results in a spin alignment of the electrons in the metal to compensate its local magnetic moment. Such an antiferromagnetic screening leads to the formation of a new many-body ground state. Due to the competition between thermal and quantum spin-flip fluctuation, however, electrons in a non-localized metallic band are not capable of screening a local magnetic moment at high temperature. Upon decreasing temperature below a crossover temperature $T_{\rm K}$ from phonon-scattering to incoherent spin-flip scattering, incoherent Kondo resonance states originating from a local magnetic moment start to emerge when the metallic electrons are gradually screening the local moment to form a Kondo singlet~\cite{Jang}.  Such screening requires, in the electron band structure point of view, the hybridization between a local electronic state of the impurity and a metallic band, resulting in the kink-like structure at their crossing point as shown in Fig.~1~\cite{YbAl3, Im, Jang}. At even lower temperature, e.\,g.\,, especially below a characteristic temperature $T_{\rm coh}$, the local magnetic moment is fully screened by conduction electrons and starts to correlate with each other via the conduction electrons to form a coherent Kondo band state, leading to an opening of a hybridization gap~\cite{Shim2}. As a result, the many-body ground state shows a strong $T$-dependence~\cite{Ternes}, leading to a characteristic resistivity versus temperature curve in transport measurements~\cite{Onuki} and temperature-dependent energy spectra near $E_{\rm F}$, Kondo resonance, and its hybridizations with a metallic band in photoemission measurements~\cite{Ehm}. 
 
  \begin{figure*}
  \begin{center}
  \includegraphics[width=0.5\columnwidth]{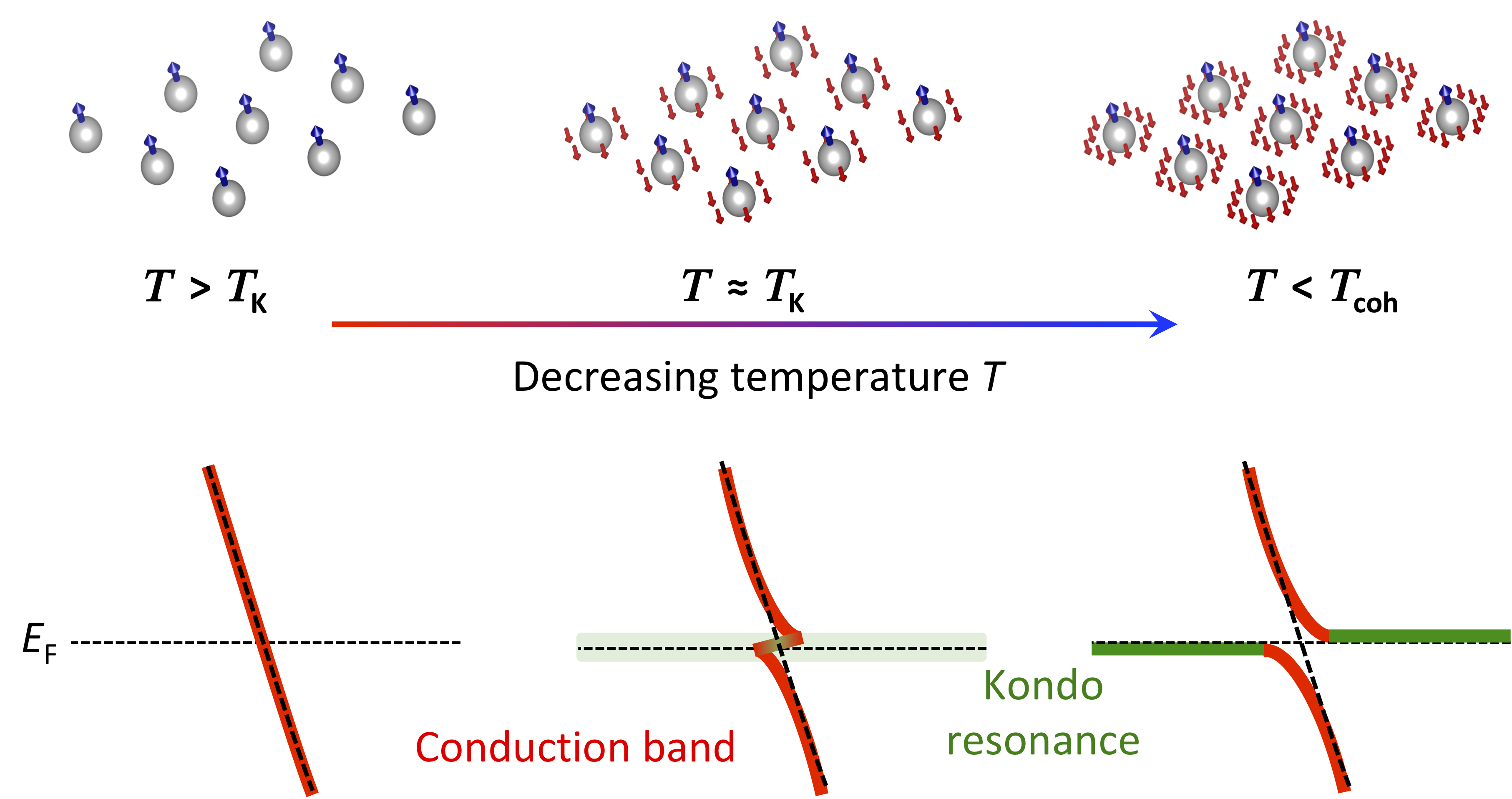}
  \end{center}
  \caption{Schematics of the spin structure and the electron band structure in Kondo regime.}
  \label{Fig1}
  \end{figure*}

The interaction between the localized states and itinerant electrons in two-dimensional (2D) systems, however, is predicted to be different from  3-dimensional (3D) cases~\cite{Zitko} and yet still far from being clearly understood in both theories and experiments~\cite{Defect,Ripple,Barua,Fritz2}. Graphene is an ideal testing ground to explore the Kondo effect in a 2D limit. Due to the Dirac fermions confined in a 2D system, exotic electronic and magnetic properties emerge which can be easily modified by external factors such as substrates~\cite{Dielectric, Pt, STO} and foreign atoms~\cite{Alkali,S}. On the other hand, vanishing charge carrier density of charge neutral graphene hinders to realize the Kondo effect as it strongly depends on the charge carrier density to screen the local magnetic moment of the impurity~\cite{Defect}. The departure from the charge neutrality, however, is predicted to bring about the Kondo effect~\cite{Vojta, Uchoa2011, Li}, which has been examined by transport~\cite{Defect} and local probe measurements~\cite{Ripple}, albeit controversial due to different behaviors depending on the impurity position and substrate~\cite{Defect2, Withoff, Donati, EelboPRB,CoSiO1,CoSiO2,CoPt}. Hence, the Kondo resonance in graphene has not been well revealed despite the intense study for the last decade. 


In order to explore the Kondo resonance in graphene, cerium atoms with a single 4$f$ electron per atom were intercalated underneath graphene grown on an SiC(0001) substrate. Figure~2(a) shows an ARPES intensity map taken across the Brillouin zone corner (K point) perpendicular to the $\Gamma$K direction of the graphene unit cell using 80 eV photons. Near Fermi energy, $E_{\rm F}$, two different branches of bands with relatively strong and weak photoelectron intensity are observed. This is similar to the ARPES data for rubidium, cesium, ytterbium, and lithium-intercalated graphene on an SiC(0001) substrate~\cite{WatcharinyanonRb, HwangYb, Caffrey}. The observation of the two spectral features has been attributed to the inhomogeneous intercalation of foreign atoms~\cite{HwangYb} or the translation of upper most graphene layer to form AA stacking induced by the intercalation~\cite{Caffrey}. The second derivative of the ARPES intensity map shown in Fig.~2(b) allows us to resolve these spectral features. For the band structure with strong spectral intensity (denoted by a grey arrow), the Dirac energy, $E_{\rm D}$, where upper and lower cones merge is $\sim$0.4~eV below $E_{\rm F}$, consistent with previous results not only for as-grown graphene on SiC(0001)~\cite{In-gap}, but also for foreign atom-intercalated graphene~\cite{WatcharinyanonRb, HwangYb, Caffrey}. For the band structure with weak spectral intensity (denoted by a purple arrow), $E_{\rm D}$ is $\sim$1.6~eV below $E_{\rm F}$. At its crossing points with the band structure with strong spectral intensity (denoted by a black arrow), the band structure is slightly modified from the characteristic linearity of the graphene $\pi$ band. Such discontinuity is reminiscent of the hybridization between the graphene $\pi$ bands with and without foreign atoms~\cite{HwangYb}. In addition, recent scanning tunneling microscopy studies reveal that intercalated manganese atoms underneath a single sheet of graphene form atomic islands preventing rest of the graphene sheet from contacting directly to the intercalated atoms~\cite{Mn} consistent with the inhomogeneous intercalation picture~\cite{HwangYb}. Indeed, the observed electron band structure is seemingly in agreement with the overlay of two tight-binding bands whose $E_{\rm D}$ is shifted towards lower energy by 0.4~eV and 1.6~eV corresponding to as-grown graphene and heavily electron-doped graphene by cerium, respectively. Here the conduction and valence bands are separated by 0.2~eV consistent with experimental results~\cite{In-gap}.

  \begin{figure*}
  \begin{center}
  \includegraphics[width=0.95\columnwidth]{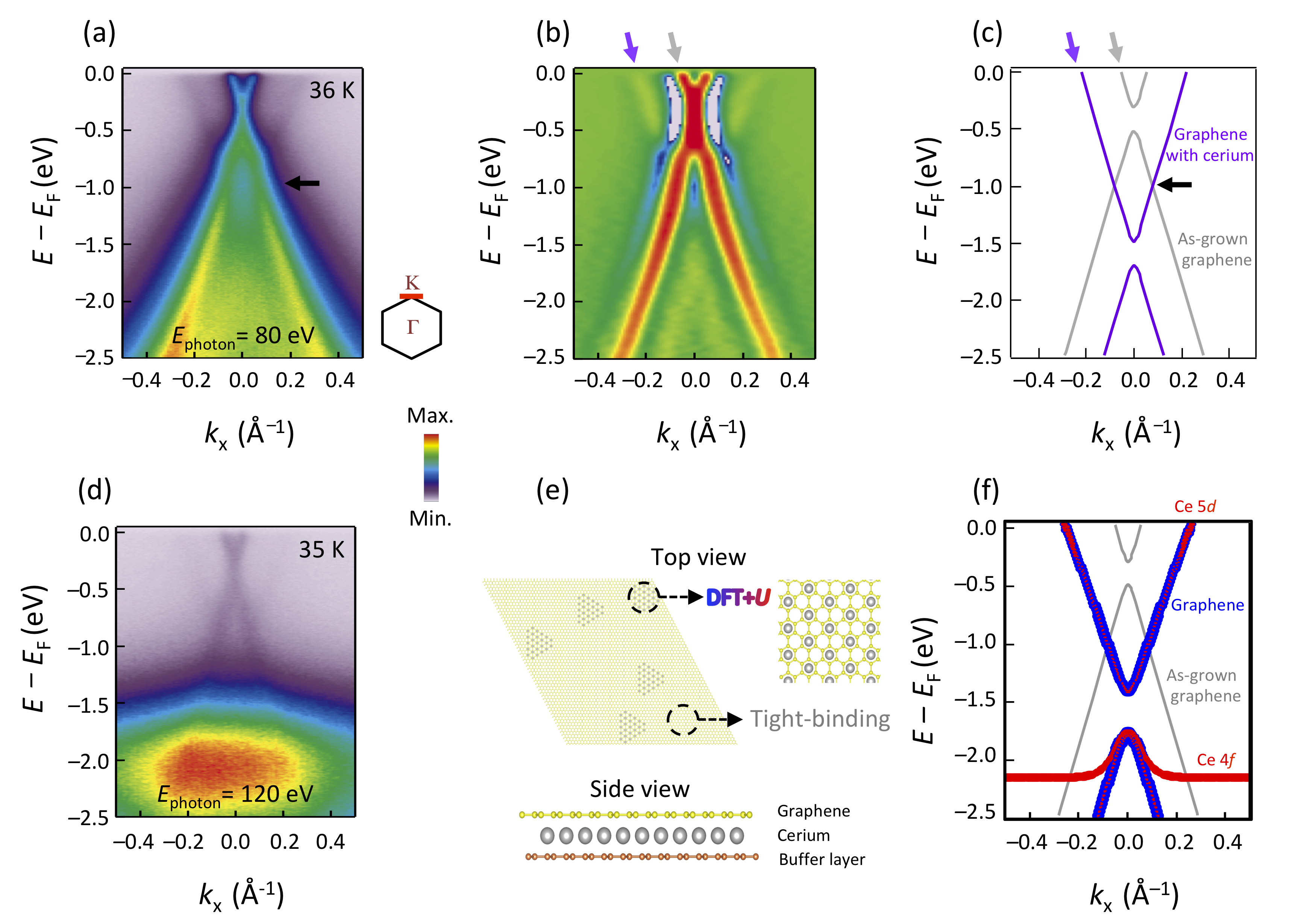}
  \end{center}
  \caption{(a, b) An ARPES intensity map (a) and its second derivative (b) for Ce-intercalated graphene taken at 36~K using 80~eV photons across the K point perpendicular to the $\Gamma$K direction of the graphene unit cell denoted by the red line in the inset. (c) Tight-binding bands perpendicular to the $\Gamma$K direction with shifted Dirac energies corresponding to as-grown graphene (grey bands) and graphene with Ce (purple bands). (d) An ARPES intensity map for Ce-intercalated graphene taken at 35~K  using 120~eV photons across the K point perpendicular to the $\Gamma$K direction. (e, f) A model structure for Ce-intercalated graphene (e) and a calculated band structure (f) perpendicular to the $\Gamma$K direction with $U=4~{\rm eV}$ and $J=0.68~{\rm eV}$. The blue and red colors denote the electronic contribution from C and Ce orbitals, respectively. The grey bands are tight-binding bands for as-grown graphene.
} 
  \label{Fig2}
  \end{figure*}


The existence of cerium atoms can be proved when using a photon energy close to the resonance photon energy for cerium~\cite{PE122}. An ARPES intensity map taken for the same sample using 120~eV photons shows not only the previously observed ARPES intensity for as-grown graphene and heavily electron-doped graphene by cerium with 80~eV photons, but also a non-dispersive state at $\sim$2.1 eV below $E_{\rm F}$. Relative change of the spectral intensity of the former stems from different scattering cross section of photons for carbon and cerium that is a function of incident photon energy~\cite{Henke}. The cross section for carbon decreases with increasing photon energy from 80~eV to 120~eV, whereas that for cerium shows a sharp resonance condition at 122~eV. To identify the non-dispersive state, band structure calculations within the density functional theory (DFT) have been performed for the geometric structure shown in Fig.~2(e) with on-site Coulomb interaction $U$~=~4~eV and exchange interaction $J$~=~0.68~eV. In Fig.~2(f), blue and red colors denote the electronic contribution from carbon 2$p_{\rm z}$ orbital and cerium orbitals, respectively. Here, the position of $E_{\rm D}$ is $\sim$1.6~eV below $E_{\rm F}$, consistent with the experimental result for weak spectral intensity. The overlay of a tight-binding band for as-grown graphene and the DFT+$U$ band for cerium-intercalated graphene shows a good agreement with experimental data shown in Fig.~2(d). In addition, close to $E_{\rm F}$, the cerium 5$d$ orbital character is observed in conjunction with the carbon 2$p_{\rm z}$ orbital character, indicating that the shift of $E_{\rm D}$ originates from the charge transfer from cerium 5$d$ to carbon 2$p_{\rm z}$ orbitals. More importantly, the comparison indicates that the non-dispersive state originates from pure charge excitations of the trivalent cerium ion (4$f^{\rm 1}$~$\rightarrow$~4$f^{\rm 0}$) that is usually referred to as the ionization peak~\cite{Ionization}, providing an evidence of the presence of cerium in this system.

 
  \begin{figure*}
  \begin{center}
  \includegraphics[width=0.95\columnwidth]{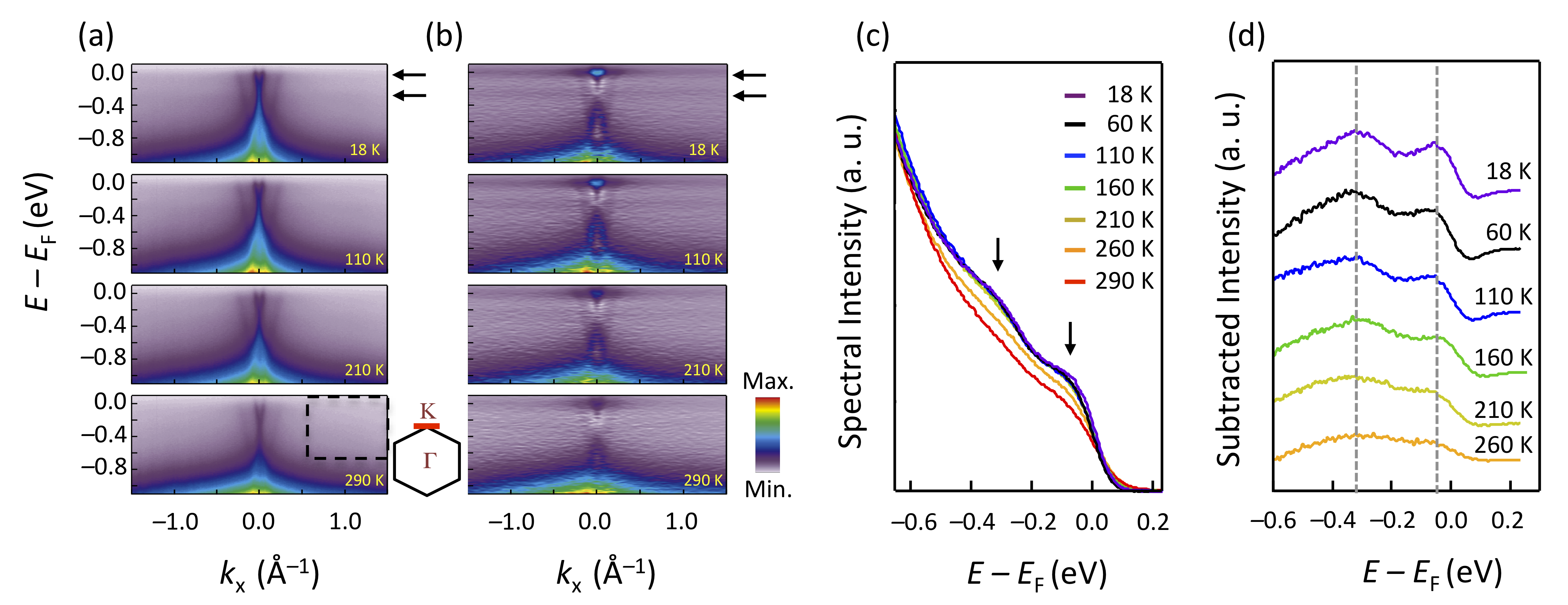}
  \end{center}
  \caption{(a, b) An ARPES intensity map (a) and its first derivative (b) for Ce-intercalated graphene taken at several different temperatures perpendicular to the $\Gamma$K direction using 122~eV photons. The intensity is normalized with respect to the highest photoelectron intensity of the area of interest. The black arrows denote two non-dispersive states that show temperature-dependence. (c) Temperature-dependent $k$-integrated energy spectra extracted from $k_{\rm x}=0.5$~\AA$^{\rm -1}$ to 1.5~\AA$^{\rm -1}$ of Fig.~3(a). The black arrows denote two humps that gradually develop with decreasing temperature. (d) Energy spectra at different temperatures after subtracting 290~K spectrum.}
  \label{Fig3}
  \end{figure*}

The temperature-dependent study on cerium-intercalated graphene shows a unique property of the interaction between graphene and cerium. Figures 3(a) and 3(b) show ARPES intensity maps of cerium-intercalated graphene and their first derivatives taken at several different temperatures ranging from 18~K to 290~K using 122~eV photons, a resonance photon energy for cerium. Here, the intensity is normalized with respect to the highest photoelectron intensity at the bottom of each panel that comes from the strong 4$f^{\rm 1}$~$\rightarrow$~4$f^{\rm 0}$ ionization peak observed at 2.1 eV below $E_{\rm F}$ shown in Fig.~2(d). Interestingly, spectral intensity near $E_{\rm F}$ exhibits temperature-dependent change. At the lowest temperature, two non-dispersive states are observed at $E_{\rm F}$ and 0.3~eV below $E_{\rm F}$ as denoted by arrows. These spectral features are not observed at an off-resonance photon energy of 80~eV shown in Fig.~2(a), indicating that they are closely correlated with cerium. Upon increasing temperature, two states gradually decrease in spectral intensity and become almost featureless at 290~K. To investigate the temperature-dependent spectral change of the two states in detail, the ARPES intensity is integrated over $k_{\rm x}$\,=\,0.5\,$\sim$\,1.5~\AA$^{-1}$ as denoted by a black dashed rectangle not to include the graphene $\pi$ band showing relatively high spectral intensity. The result is summarized in Fig.~3(c). From the featureless spectrum taken at 290 K, two humps are gradually developed with decreasing temperature, as denoted by black arrows, in conjunction with the increasing slope of the spectrum at $E_{\rm F}$ mainly due to the reduced thermal excitation of the Fermi-Dirac distribution~\cite{FDD}. When the 290~K spectrum is subtracted as shown in Fig.~3(d), the energy spectra clearly shows the evolution of the two peaks at 0.05~eV and 0.33~eV below $E_{\rm F}$ especially below 110~K $\sim$ 160~K.
 
One of the plausible scenarios explaining the observed temperature-dependent evolution of non-dispersive states near $E_{\rm F}$ is the  Kondo effect~\cite{Patthey, Ehm}. The ionization peak at higher energy (2.1~eV below $E_{\rm F}$) represents the localized cerium 4$f$ electron that plays a role of a local magnetic moment. When temperature is sufficiently low, the local magnetic moment can be antiferromagnetically screened by the spins of surrounding metallic electrons from heavily electron-doped graphene by cerium, possibly leading to the formation of a hybridized many-body ground state. Theoretically, such a Kondo effect can be examined when the electronic correlations are fully taken into account via the DFT combined with the dynamic mean field theory (DFT+DMFT)~\cite{Shim1, Shim2, CeIn3, Kang}. Figures 4(a) and 4(b) show the calculated electron band structure for the cerium-intercalated graphene for the same model used for DFT+$U$ calculations using the DFT+DMFT approach for 300~K and 20~K, respectively. The band structure calculated for~300 K shows the graphene $\pi$ band that is electron-doped by cerium 5$d$ electrons as shown in Fig.~4(a), consistent with the result obtained using the DFT+$U$ approach shown in Fig.~2(f). 
In the calculated band structure for 20~K shown in Fig.~4(b), the most notable change is the emergence of well-defined two non-dispersive states at $E_{\rm F}$ and 0.31~eV below $E_{\rm F}$, consistent with experimental results shown in Fig.~3, which are the Kondo resonance and the spin-orbit side band, respectively. 

The emergence of the two states at low temperature in both experiments and calculations manifests itself the hybridization between a cerium 4$f$ state and the heavily electron-doped graphene $\pi$ band by cerium to form a new resonance many-body ground state with quenched magnetic moment~\cite{Jang, Ehm}. Such a hybridization requires a kink-like structure in the metallic band at the energies for the Kondo resonance and the spin-orbit side band as schematically shown in Fig.~1~\cite{Shim2}. Indeed, the calculated band structure for 20~K shows moderate kinks in the metallic band at the crossing points with the Kondo resonance and the spin-orbit side band as shown in Fig.~4(b), although it does not show a fully opened hybridization gap. Note that such kink-like structure is not predicted in the calculated electron band structure for 300~K as shown in Fig.~4(a). To experimentally prove the presence of the kink-like structure, the ARPES intensity map for heavily electron-doped graphene by cerium is analyzed by a Lorentzian fit to $\sim$200 momentum distribution curves (MDCs) as shown in Fig.~4(c), a few of them shown on the right panel. The resultant red curve shows a kink-like structure at $\sim$0.33~eV below $E_{\rm F}$ (denoted by an arrow), the energy of the spin-orbit side band, in agreement with the DMFT results, while the kink-like structure for Kondo resonance at $E_{\rm F}$ is not observed due to the Fermi-Dirac distribution.

  \begin{figure*}
  \begin{center}
  \includegraphics[width=0.5\columnwidth]{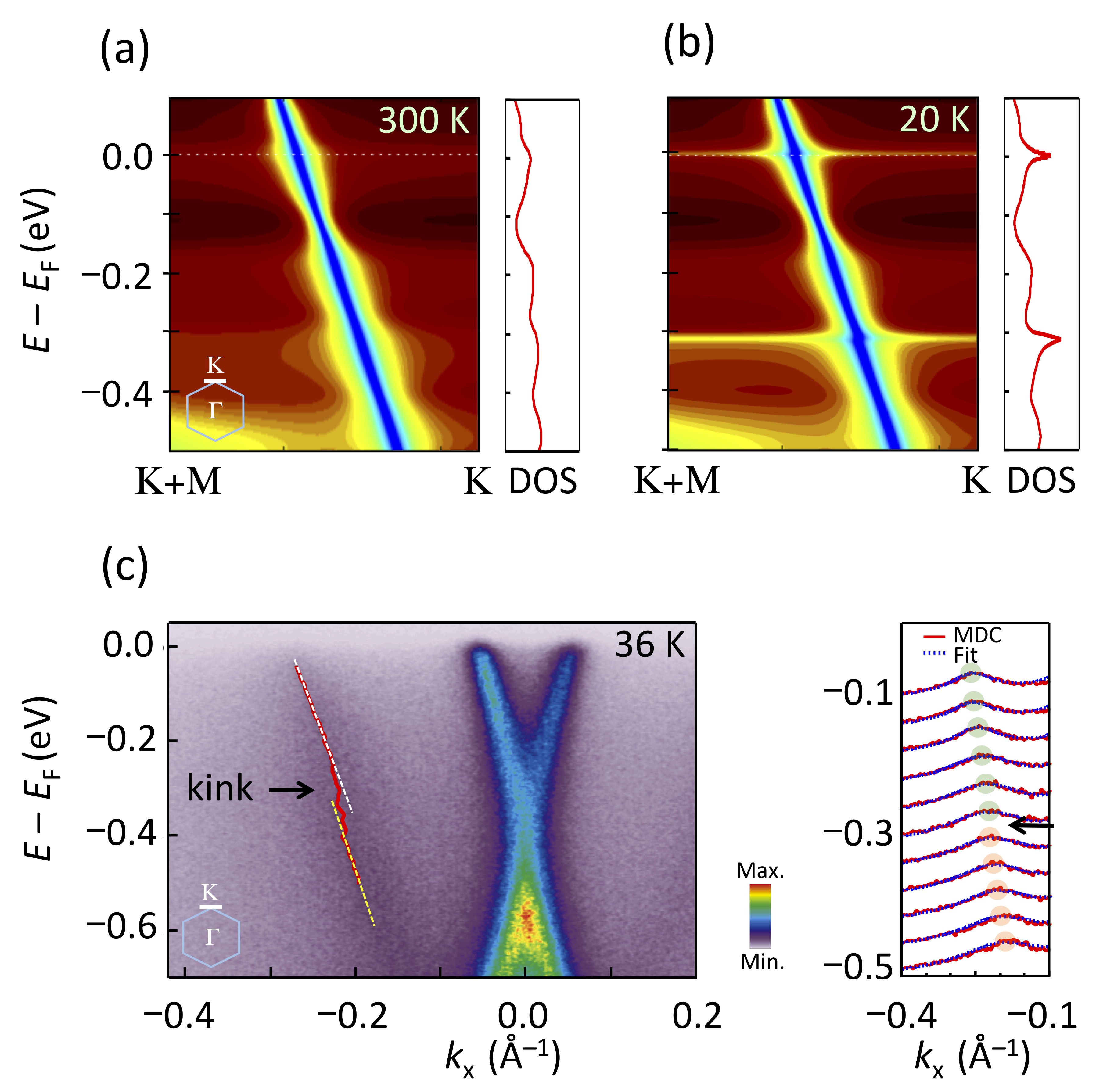}
  \end{center}
  \caption{(Color online) (a, b) The calculated electron band structure within the DFT+DMFT approach at (a) 300~K and (b) 20~K . The right panel is a density of states (DOS) spectrum for each case. (c) An ARPES intensity map near $E_{\rm F}$ taken at 36~K with 80~eV photons. The red line is an energy-momentum dispersion obtained using a Lorentzian fit to $\sim$200 momentum distribution curves for Ce-intercalated graphene, when the right panel shows a few of them (red curves) with their Lorentzian fit (blue dotted curves). The white and yellow dashed lines in the left panel are arbitrary straight lines for guide to the eyes.  The pale green and orange circles in the right panel denote the maximum point of the momentum-distribution curves.}
  \label{Fig4}
  \end{figure*}


The observed temperature dependence provides important information on the Kondo physics in graphene with external magnetic impurities. Within the Kondo picture, there exists  two temperature scales, i.\,e.\,, the temperature where metallic electrons start to screen a local magnetic moment, so-called Kondo temperature $T_{\rm K}$, and the temperature where the local magnetic moment starts to correlate with each other via the metallic electrons, so-called coherent temperature $T_{\rm coh}$. In the energy spectrum obtained after subtracting a featureless background measured at 290 K from the spectrum taken at several different temperatures shown in Fig. 3(d), Kondo resonance and the spin-orbit side band at $E_{\rm F}$ and 0.3 eV below $E_{\rm F}$ start to emerge below 110~K $\sim$ 160~K, providing the temperature scale of $T_{\rm K}$. At the lowest experimental temperature, however, the spectral intensity of Kondo resonance and the spin-orbit side band, and the kink in the electron band structure are still very weak compared to the signal observed from typical Kondo systems~\cite{Jang,Allen}. It is important to note that even the weak spectral intensity and the moderate kink are in good agreement with DMFT calculations, suggesting that the weak signal does not imply low quality of the experimental data. Instead, the agreement indicates that the coherent Kondo band state is not fully developed even at the lowest experimental temperature that we can access, i.\,e.\,, $T_{\rm coh}<$\,20~K. 

The low $T_{\rm coh}$ is attributed to the intrinsic Kondo physics in graphene with external magnetic impurities. The reasons are as follows. First, longer inter-cerium distance can lead to lower $T_{coh}$. Note that the inter-cerium distance of $\alpha$- and $\gamma$-phase cerium is 3.411~\AA~and  3.649~\AA, respectively~\cite{Distance}, and that $\gamma$-phase shows an order of magnitude weaker spectral intensity of Kondo resonance~\cite{CeGA}. For cerium-intercalated graphene shown in Fig.~2(e), inter-cerium distance is 4.270~\AA, even longer than that of $\gamma$-phase cerium, resulting in even lower spectral intensity of Kondo resonance. As a result, the coherent band state between local Kondo singlet states is not observed. 
The characteristics of epitaxial graphene on an SiC substrate can influence the development of Kondo resonance as well. The coexistence of strong spectral intensity for as-grown graphene and weak spectral intensity for heavily electron-doped graphene by cerium suggests that absolute concentration of cerium is small. In addition, the lateral size of typical epitaxial graphene on an SiC substrate ranges from 50~nm to 200~nm~\cite{David}, while intercalated atomic islands can cover only a few tens nm~\cite{Mn}. As a result, there might be a finite size effect that can compete with the Kondo effect by lifting up the degeneracy of the metallic state. Indeed, quantum dots with a tunable number of electrons exhibit the Kondo effect at temperature lower than 1~K~\cite{Cronenwett}.


In summary, through a combined study of ARPES and DMFT, we provide evidence for the 4$f$ induced Kondo resonance and the spin-orbit side band, suggesting that the Kondo effect can be realized in graphene in the presence of external magnetic impurities~\cite{Impurity}. The Kondo physics in graphene will shed light on the fundamental understanding on many-body physics in a prototypical 2D system with Dirac electrons, and provide a versatile route towards the device applications such as graphene-based single-electron transistor~\cite{Transistor} and spintronics devices~\cite{Spintronics}. Since graphene is inert and intercalants underneath graphene can be protected from oxidations, we expect that metal-intercalated graphene should be an ideal playground to understand and utilize unique many-body physics that appear in a two-dimensional limit.

{\bf Experimental Section.} {\em Sample preparation and ARPES.} Single-layer graphene samples were epitaxially grown on a 6$H$-SiC(0001) substrate under Si flux in a high vacuum chamber with a base pressure of $3\times10^{-8}$~Torr~\cite{HwangYb}. The cerium was intercalated underneath the graphene sample by evaporating from a tungsten coil wrapping ultrahigh purity cerium (99.9\%) that is followed by annealing at 840~$^{\circ}{\rm C}$. The cerium-intercalated graphene was transferred to an ultra-high vacuum chamber for ARPES measurements. Before each measurement, the sample was heated upto 760~$^{\circ}{\rm C}$ to remove contaminations in a sample preparation chamber connected to the ARPES chamber. ARPES measurements were performed at Beamlines 10.0.1 and 4.0.3, Advanced Light Source, Lawrence Berkeley National Laboratory. ARPES data were obtained with a Scienta R4000 electron analyzer at several different temperatures ranging from 290~K to 18~K. The energy and angular resolutions are 56~meV (determined at 24~K) and 0.1$^{\circ}$.

{\em Band structure calculation.} The electronic band structure calculations have been performed within the density functional theory by using the full-potential local-orbital (FPLO) band method~\cite{FPLO}. The unit cell we have considered contains one Ce atom on top of the hollow site of $\sqrt{3}\times\sqrt{3}$ graphene unit cell. The atomic positions are relaxed by using the Hellman-Feynman force theorem. The resulting structure resembles a single layer CeC$_6$ structure. To describe the 4$f$ state of Ce, density-functional theory (DFT)$+U$ calculation has been done for the same geometric structure with on-site Coulomb interaction $U=4~\rm{eV}$ and exchange interaction $J=0.63~\rm{eV}$. The $U$ parameter was chosen so that the position of lower Hubbard band agrees reasonably with that of ARPES results. The DFT+dynamic mean field theory (DMFT) calculation implemented in the Wien2k code has utilized for the realistic description of many-body effects and the temperature evolution of the electron band structure of Ce-intercalated graphene~\cite{CeGA, Wien2k}. One crossing approximation~\cite{OCA1, OCA2} is used for the impurity solver. $U$ and $J$ are chosen to be 6~eV and 0.68~eV, respectively, for the DFT+DMFT calculation.

\begin{acknowledgement}
This work was supported by the National Research Foundation of Korea (NRF) grant funded by the Korea government (MSIP) (No.~2015R1C1A1A01053065, No.~2017K1A3A7A09016384, and No.~2018R1A2B6004538). JH acknowledges support from NRF-2017-Fostering Core Leaders of the Future Basic Science Program/Global Ph.D. Fellowship Program. The work in Max Planck POSTECH Center for Complex Phase Materials was supported by the National Research Foundation of Korea (NRF) funded by the Ministry of Science, ICT and Future Planning (No. 2016K1A4A4A01922028). The Advanced Light Source is supported by the Office of Basic Energy Sciences of the U.S. Department of Energy under Contract No. DE-AC02-05CH11231. The work in University of California, Berkeley, was supported by Berkeley Lab's program on sp2 bond materials, funded by the U.S. Department of Energy, Office of Science, Office of Basic Energy Sciences, Materials Sciences and Engineering Division, of the U.S. Department of Energy (DOE) under Contract No.~DE-AC02-05CH11231. KK acknowledges support from the National Research Foundation of Korea (NRF) grant funded by the Korea government (MSIP) (NRF-2016R1D1A1B02008461 and NRF-2017M2A2A6A01071297). BIM acknowledges the support from the NRF (Contracts No.  2017R1A2B4005175). JWC acknowledges support from the National Research Foundation of Korea (NRF) grant funded by the Korea government (NRF-2015R1A5A1009962). Experiments at PLS-II were supported in part by MSIP and POSTECH.
\end{acknowledgement}

\end{document}